\PassOptionsToPackage{nonamebreak}{natbib}
\PassOptionsToPackage{desactivate}{linenoaa}
\def\linenumbers{}
\def\nolinenumbers{}
\def\columnwiselinenumbersfalse{}
\def\runningpagewiselinenumbers{}

\documentclass{aa}

\usepackage{graphicx}
\usepackage{txfonts}
\usepackage{lipsum,xcolor}
\usepackage{subcaption}
\usepackage{lscape}
\usepackage{placeins}
\usepackage{tikz}
\usetikzlibrary{arrows.meta,positioning}

\usepackage[colorlinks=true,linkcolor=blue,citecolor=blue,urlcolor=blue]{hyperref}
\usepackage{orcidlink}
\makeatletter
\def\hyper@natlinkbreak#1#2{#1}
\renewcommand*\aa@manuscriptname{manuscript no. aa63717-26\hspace{\stretch{1}}\copyright ESO \the\year}
\renewcommand*\aa@textidlineempty{{\slshape A\&A proofs:}\ manuscript no.~aa63717-26}
\makeatother

\begin{document}

   \title{Optimal cosmological cross-correlation analysis\\ with U-Net-based processing}

   \author{W. Yan\orcidlink{0009-0005-5169-9741}\inst{1}\corrauth{yanweitong449@gmail.com}
        \and D. Gruen\orcidlink{0000-0003-3270-7644}\inst{1,2}
        \and O. Friedrich\orcidlink{0000-0001-6120-4988}\inst{1,2}
        \and A. Kov\'acs\orcidlink{0000-0002-5825-579X}\inst{3,4}
        }

   \institute{University Observatory, Faculty of Physics, Ludwig-Maximilians-Universität, Scheinerstr. 1, 81679 Munich, Germany \and Excellence Cluster ORIGINS, Boltzmannstr. 2, 85748 Garching, Germany \and MTA-CSFK Lendület Large-Scale Structure Research Group, Konkoly-Thege Miklós út 15-17, H-1121 Budapest, Hungary \and Konkoly Observatory, HUN-REN CSFK, MTA Centre of Excellence, Budapest, Konkoly Thege Miklós út 15-17. H-1121, Hungary}

   \abstract
   {Cosmological cross-correlation analyses are commonly based on two-point statistics, which are optimal for Gaussian random fields but do not capture all relevant information in the presence of non-Gaussianity, which is ubiquitous in late-time cosmic fields. This motivates investigations of higher order statistics, and one method is to use machine learning to construct a correlation-optimized tracer field that can utilize cross-correlation information beyond two-point statistics.}
   {Using a tracer field as input and a paired target field as part of the cost function, we test whether a U-Net can learn a transformed tracer field that maximally correlates with the target field.}
   {We train a 1D U-Net on paired simulated fields and study controlled cases based on Gaussian and log-normal random fields. We train and quantify performance using the Pearson correlation coefficient with the target field and compare the network output with the analytic linear solution known in the Gaussian limit.}
   {For Gaussian fields, the U-Net reproduces the expected linear solution closely and behaves as an effective matched filter in frequency space. For log-normal fields, the network departs from the linear estimator and achieves an improved correlation with the target field, increasing the mean Pearson correlation from $0.857$ for the linear estimator to $0.953$ for the U-Net output. This indicates that the network is sensitive to information beyond the power spectrum.}
   {These controlled 1D experiments show that U-Net-based tracer transformation can effectively extract correlation-relevant information beyond the power spectrum, providing a promising framework for enhancing cosmological cross-correlation analyses.}


   \maketitle

\section{Introduction}
Modern cosmology increasingly relies on statistical inference from the cosmic microwave background (CMB) and large-scale structure (LSS), and a key challenge is to extract maximal information from these complex fields.
In recent years, machine learning methods, and in particular Convolutional Neural Networks (CNNs) \citep{LeCun_2015}, have been applied to a wide range of cosmological inference tasks. These include parameter estimation and model discrimination from galaxy density and weak gravitational lensing fields \citep[e.g.][]{Gupta_2018, Fluri_2018, Kacprzak_2022, Lu_2023, Gong_2024, Jeffrey_2025}, 21-cm tomography during cosmic dawn and reionization \citep[e.g.][]{Gillet_2019}, strong gravitational lens modeling \citep[e.g.][]{PerreaultLevasseur_2017}, analysis of the 1D Lyman-$\alpha$ forest \citep[e.g.][]{Nayak_2024,Chang_2025}. CNN-based analyses have also been used for multifield cosmological inference from hydrodynamical simulations \citep{VillaescusaNavarro_2021}.
Among these architectures, the U-Net \citep{ronneberger2015} is especially well suited to map-to-map regression problems. Its encoder-decoder structure, together with skip connections, enables the network to combine local information with multi-scale contextual features while preserving spatial resolution in the output.
U-Nets have therefore been successfully applied in various cosmological analyses, including weak-lensing map reconstruction \citep{Jeffrey_2020}, mitigation of polarization leakage in neutral hydrogen mapping \citep{Gao_2023}, and Cosmic Microwave Background (CMB) foreground cleaning \citep{Petroff_2020}.

Another prospective application of the U-Net that has not previously been explored is to extract non-Gaussian information from large-scale structure (LSS) in order to optimize cross-correlation analyses.
Traditional cosmological cross-correlation studies usually rely on 2-point statistics.
The 2-point correlation function (2PCF), or its Fourier counterpart the power spectrum, is the most widely used summary statistic in observational cosmology,
including tomographic cosmic shear analyses (DES: \citealt{Amon_2022}; HSC: \citealt{Li_2023}; KiDS: \citealt{Wright_2025}),
galaxy clustering and baryon acoustic oscillation measurements (DES: \citealt{Alam_2021}; DESI: \citealt{DESI_2025}),
full-shape galaxy power-spectrum analyses (SDSS: \citealt{Brieden_2023}; DESI: \citealt{DESI_2025_FS}),
and CMB lensing (Planck: \citealt{Carron_2022}; ACT: \citealt{Madhavacheril_2024}; SPT: \citealt{Ge_2025}).

The initial density fluctuations of the Universe are highly Gaussian, as is reflected e.g.~in the highly Gaussian nature of CMB temperature fluctuations. In such a case, all relevant information is encoded in the power spectrum.
However, late time non-linear gravitational evolution induces non-Gaussianities, where higher-order statistics also contain relevant information.
Accessing this information could in principle improve the detectability of weak cross-correlation signals, and increase the stringency of cosmological tests. However, higher-order statistics are often computationally expensive to measure and predict and their likelihood difficult to interpret, and are therefore less widely used in practice, even more so for signals of intrinsically low significance.

More generally, a long-standing challenge in LSS analyses is to identify manageable summary statistics that are as close as possible to sufficient statistics, in the sense of retaining most of the cosmological information present in a non-Gaussian field \citep{Carron_2013}. Examples include the log power spectrum, which recovers information beyond the standard power spectrum on translinear scales \citep{Neyrinck_2009, Repp2016}, and marked power spectra, which reweight the field with certain marked functions to access beyond-two-point information within a familiar analysis framework \citep{EbinaWhite2025, Cowell_2025}.
The method pursued here follows the same general aim, but instead of specifying the transformation by hand, we train a neural network to learn to produce a correlation-optimized tracer field from paired input fields, one as input to the network and the other used only in the loss function.

This motivates a U-Net-based approach to learning a correlation-optimized tracer field from two fields, that is, a  transformation of the input field where the transformation parameters depending on the local environment of the input field are learned by CNNs, so that the transformed field is optimized to correlate with the target field.
The U-Net architecture is well suited to this map-to-map learning task because it captures multi-scale features without imposing an explicit functional form.
The goal of this paper is therefore to test whether a U-Net can learn a correlation-optimized tracer field that captures information beyond standard 2-point statistics and thereby improves the signal-to-noise ratio of downstream cross-correlation analyses.
This can be subsequently applied to various cosmological cross-correlation scenarios, such as the late-time Integrated Sachs-Wolfe (ISW) effect, which relies on cross-correlating the primary CMB anisotropies with large-scale structure (LSS) tracers \citep[e.g.][]{Planck_2016_ISW, Dong_2022}, or CMB lensing tomography, which cross-correlates the reconstructed CMB lensing convergence with galaxy overdensity fields to constrain the amplitude and growth of structure \citep[e.g.][]{Krolewski_2021, Farren_2024}.

To assess this idea in a controlled way, we first study simplified 1D test cases based on Gaussian and log-normal random fields. In machine learning applications, it is often difficult to determine whether an acceptable, a superior, or an optimal solution has been achieved. Here, however, we can study controlled scenarios that allow to evaluate whether the U-Net based 2-point approach can indeed extract the full cross-correlation information.
The Gaussian case provides a useful benchmark, since the expected behaviour is analytically tractable, while the log-normal case introduces a more realistic form of non-Gaussianity commonly encountered in LSS applications.

The paper is organized as follows. In Section \ref{chap:unet}, we introduce the U-Net architecture used for this correlation-optimization problem. In Section \ref{chap:1d}, we apply the U-Net to 1D toy-model test cases involving Gaussian and log-normal random fields.
Finally, in Section \ref{chap:conclusion}, we summarize our findings and discuss directions for future research.

\section{U-Net based correlation optimization} \label{chap:unet}
In this section, we introduce the methodology used to learn correlation-optimized tracer fields. The goal is to construct, from an input tracer field, a transformed tracer field that is maximally correlated with a target field of interest.
In the notation used throughout this paper, we denote the input tracer field by \textbf{I}, the target field by \textbf{T}, and the network output by \textbf{O}. In addition, a linear Gaussian estimator \textbf{L} is defined as the result of applying a linear frequency filter to \textbf{I} which maximizes the correlation between \textbf{L} and \textbf{T} in the limit where \textbf{I} and \textbf{T} are correlated Gaussian random fields.

The role of the network is therefore to learn a mapping
\begin{equation}
    \textbf{O} = f_{\boldsymbol{\theta}}(\textbf{I}),
\end{equation}
where the parameters $\boldsymbol{\theta}$ are optimized such that \textbf{O} correlates as strongly as possible with \textbf{T}.

The objective is to learn a transformed tracer field that is better suited for downstream cross-correlation analyses. In particular, if the relevant fields are jointly Gaussian, the optimal estimator is expected to be linear, so that \textbf{O} = \textbf{L}. The ideally trained network is hence effectively a frequency filter according to the ratio of auto- and cross power spectra. This Gaussian limit therefore provides an important sanity check. For non-Gaussian fields, however, a nonlinear model may exploit information beyond the power spectrum, allowing \textbf{O} to outperform \textbf{L} in its correlation with \textbf{T}.

Our methodology is based on supervised learning using pairs of simulated input and target fields. During training, the target field \textbf{T} is available only to define the loss function, while the network acts only on the input field \textbf{I}. This setup is designed to test whether a learned nonlinear transformation of the tracer field can extract correlation-relevant information that is not captured by the power spectrum alone.

The U-Net is a natural choice for this task because it is a multi-scale map-to-map architecture that combines local and large-scale information while preserving spatial resolution. Its encoder-decoder structure allows the network to identify broad contextual features, and its skip connections retain localized information that may otherwise be lost during downsampling.

\textbf{O} can be interpreted as U-net's attempt to reconstruct the target field \textbf{T}, and this correlation-optimized tracer field is intended for downstream cross-correlation analyses. The approach is related to the marked-correlation method used in cosmology studies (e.g.~\citealt{Cowell_2024, Cowell_2025}), which reweights the tracer field with a mark function in order to extract higher-order information.
The output \textbf{O} can be seen as a marked version of the input field \textbf{I}, where the mark depends on the local environment of a point within \textbf{I} and is optimized to correlate with the target field \textbf{T}.
While both methods aim to enhance signal extraction by reweighting the tracer field, our approach does not specify a mark function a priori, but instead learns a spatially varying transformation directly from the data that depends on the environment of the point, so it can adapt to the specific characteristics of the input fields.

We begin with a 1D proof-of-concept, where the behavior of the method can be tested in controlled Gaussian and log-normal settings.
\subsection{1D U-net architecture}\label{sec:1dunet}
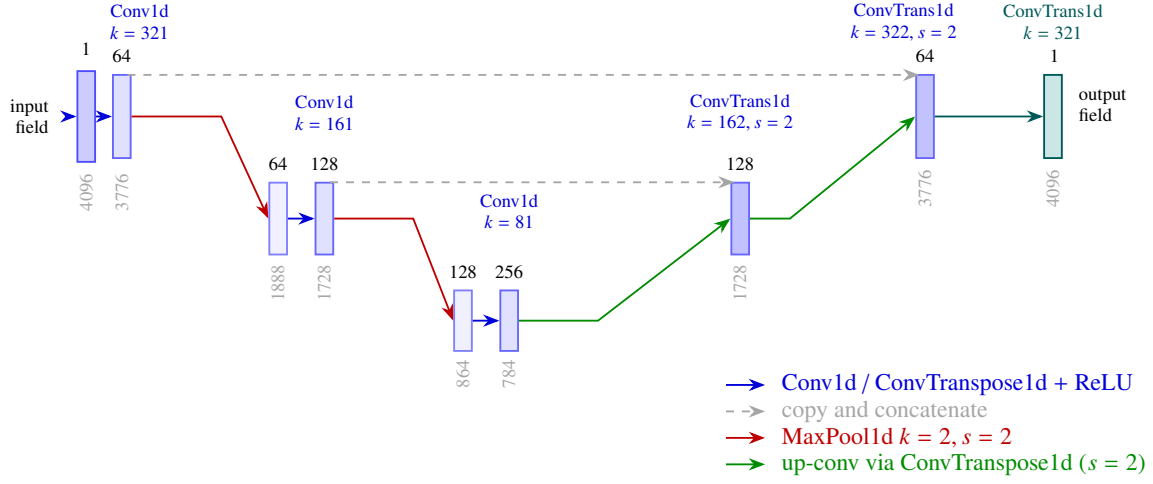
\begin{figure*}[htbp]
    \centering
        \begin{tikzpicture}[
            x=0.94cm,
            y=1cm,
            line width=0.7pt,
            font=\scriptsize,
            >=Stealth,
            enc/.style={draw=blue!60, fill=blue!12, minimum width=0.16cm},
            poolb/.style={draw=blue!50, fill=blue!6, minimum width=0.14cm},
            dec/.style={draw=blue!60, fill=blue!24, minimum width=0.16cm},
            inb/.style={draw=blue!70, fill=blue!20, minimum width=0.14cm},
            outb/.style={draw=teal!70!black, fill=teal!20, minimum width=0.14cm}
        ]

        \node[inb, minimum height=1.20cm] (in0) at (-0.80,3.80) {};
        \node[enc, minimum height=1.10cm] (e1) at (-0.30,3.80) {};
        \node[poolb, minimum height=0.95cm] (p1) at (1.90,2.45) {};
        \node[enc, minimum height=0.95cm] (e2) at (2.55,2.45) {};
        \node[poolb, minimum height=0.80cm] (p2) at (4.50,1.10) {};
        \node[enc, minimum height=0.80cm] (e3) at (5.15,1.10) {};

        \node[dec, minimum height=0.95cm] (d1) at (8.40,2.45) {};
        \node[dec, minimum height=1.10cm] (d2) at (11.00,3.80) {};
        \node[outb, minimum height=1.10cm] (d3) at (12.80,3.80) {};

        \node at (-0.80,4.70) {1};
        \node at (-0.30,4.60) {64};
        \node at (1.90,3.18) {64};
        \node at (2.55,3.18) {128};
        \node at (4.50,1.75) {128};
        \node at (5.15,1.75) {256};
        \node at (8.40,3.18) {128};
        \node at (11.00,4.60) {64};
        \node at (12.80,4.60) {$\,1$};

        \node[rotate=90, text=gray!70] at (-0.80,2.80) {$4096$};
        \node[rotate=90, text=gray!70] at (-0.30,2.80) {$3776$};
        \node[rotate=90, text=gray!70] at (1.90,1.60) {$1888$};
        \node[rotate=90, text=gray!70] at (2.55,1.60) {$1728$};
        \node[rotate=90, text=gray!70] at (4.50,0.40) {$864$};
        \node[rotate=90, text=gray!70] at (5.15,0.40) {$784$};
        \node[rotate=90, text=gray!70] at (8.40,1.60) {$1728$};
        \node[rotate=90, text=gray!70] at (11.00,2.85) {$3776$};
        \node[rotate=90, text=gray!70] at (12.80,2.85) {$4096$};

        \node[align=right] at (-1.60,3.80) {input\\field};
        \node[align=left] at (13.50,3.95) {output\\field};

        \draw[->, blue!85!black] (-1.08,3.80) -- (in0.west);
        \draw[->, blue!85!black] (in0.east) -- (e1.west);

        \coordinate (p1a) at (1.10,3.80);
        \draw[->, red!75!black] (e1.east) -- (p1a) -- (p1.west);
        \draw[->, blue!85!black] (p1.east) -- (e2.west);

        \coordinate (p2a) at (3.80,2.45);
        \draw[->, red!75!black] (e2.east) -- (p2a) -- (p2.west);
        \draw[->, blue!85!black] (p2.east) -- (e3.west);

        \coordinate (u1a) at (6.40,1.10);
        \draw[->, green!55!black] (e3.east) -- (u1a) -- (d1.west);

        \coordinate (u2a) at (9.10,2.45);
        \draw[->, green!55!black] (d1.east) -- (u2a) -- (d2.west);

        \draw[->, teal!70!black] (d2.east) -- (d3.west);

        \draw[->, dashed, gray!70] (-0.22,4.35) -- (10.92,4.35);
        \draw[->, dashed, gray!70] (2.63,2.93) -- (8.32,2.93);

        \node[text=blue!85!black, align=center] at (-0.05,5.05) {Conv1d\\$k=321$};
        \node[text=blue!85!black, align=center] at (2.55,3.85) {Conv1d\\$k=161$};
        \node[text=blue!85!black, align=center] at (5.15,2.55) {Conv1d\\$k=81$};
        \node[text=blue!85!black, align=center] at (8.40,3.85) {ConvTrans1d\\$k=162,s=2$};
        \node[text=blue!85!black, align=center] at (10.70,5.05) {ConvTrans1d\\$k=322,s=2$};
        \node[text=teal!70!black, align=center] at (12.80,5.05) {ConvTrans1d\\$k=321$};

        \draw[->, blue!85!black] (8.2,0.25) -- +(0.55,0);
        \node[anchor=west, text=blue!85!black, font=\footnotesize] at (8.85,0.25) {Conv1d / ConvTranspose1d + ReLU};
        \draw[->, dashed, gray!70] (8.2,-0.10) -- +(0.55,0);
        \node[anchor=west, text=gray!70, font=\footnotesize] at (8.85,-0.10) {copy and concatenate};
        \draw[->, red!75!black] (8.2,-0.45) -- +(0.55,0);
        \node[anchor=west, text=red!75!black, font=\footnotesize] at (8.85,-0.45) {MaxPool1d $k=2,s=2$};
        \draw[->, green!55!black] (8.2,-0.80) -- +(0.55,0);
        \node[anchor=west, text=green!55!black, font=\footnotesize] at (8.85,-0.80) {up-conv via ConvTranspose1d ($s=2$)};
    \end{tikzpicture}
    \caption{One-dimensional U-Net architecture diagram with depth=3 for an 1D input field sampled on 4096 pixels. The numbers on top of the blocks indicate the number of channels, and the numbers at the bottom indicate the length of the feature maps at each level. The arrows indicate the flow of data through the network, with different colors representing different types of operations (convolution, pooling, up-convolution, and skip connections).}
    \label{fig:unet1d}
\end{figure*}
As a proof of concept, we first test the U-Net on simplified 1D toy scenarios using Gaussian and log-normal random fields as input and target fields.
Fig.~\ref{fig:unet1d} shows the architecture of the 1D U-Net used in this work. The input is a 4096-pixel 1D field (Gaussian or log-normal), representing an LSS tracer field, while the latent target field has the same resolution and represents the ISW field. The output has the same resolution as the input and is interpreted as a transformed tracer field optimized for correlation with the target.
We use the negative Pearson correlation (normalized cross-variance) as the loss function $\mathcal{L}$, which is defined as
\begin{equation}
    \mathcal{L} = -\frac{\sum_{i=1}^N (x_i - \bar{x})(y_i - \bar{y})}{\sqrt{\sum_{i=1}^N (x_i - \bar{x})^2} \sqrt{\sum_{i=1}^N (y_i - \bar{y})^2}},
    \label{eq:loss}
\end{equation}
where $x$ is the network output and $y$ is the latent target field, and $\bar{x}$ and $\bar{y}$ are their mean values. The Pearson correlation ranges from -1 to 1, where 1 means perfect positive correlation (meaning the two fields are identical up to a coefficient), -1 means perfect negative correlation, and 0 means no correlation. By minimizing the negative Pearson correlation, we encourage the U-Net to learn features from the input fields so that the output fields are maximally correlated with the target fields.
By fine-tuning the hyperparameters of the U-net, such as the number of layers, the kernel sizes, etc, we can further improve the performance of the network. However, the current architecture is sufficient to demonstrate how the U-net can learn to extract relevant correlation information, and performs near-optimally for controlled experiments (e.g. see subsection~\ref{sec:gaussian2}), therefore we proceed with this configuration for the experiments presented in this work.

\subsection{Optimal linear estimator for Gaussian fields}
For Gaussian fields the optimal transformation can be derived analytically.
Given two real valued statistically homogeneous Gaussian random fields $\mathbf{I}$ and $\mathbf{T}$,
with auto-power spectra $P_{II}(k)$, $P_{TT}(k)$ and
cross-power spectrum $P_{IT}(k)$,
we look for the linear, translation-invariant estimator
$\mathbf{L}$ that is maximally correlated with $\mathbf{T}$.
Any such estimator is a convolution of the input and hence diagonal in Fourier space,
\begin{equation}
    \tilde{L}(k) = W(k)\,\tilde{I}(k),
\end{equation}
where tilde denotes the Fourier transform, and $W(k)$ is a filter to be determined.
By Parseval's theorem the ensemble averages in Eq.~(\ref{eq:loss}) become sums over modes,
\begin{equation}
    \begin{aligned}
        \langle L T\rangle &= \sum_k W(k)\,P_{IT}(k), \qquad
        \langle L^2\rangle = \sum_k W(k)^2\,P_{II}(k), \\
        \langle T^2\rangle &= \sum_k P_{TT}(k),
    \end{aligned}
\end{equation}
so that the correlation coefficient is a functional of the filter,
\begin{equation}
    r[W] = \frac{\sum_k W(k) P_{IT}(k)}
                {\sqrt{\sum_k W(k)^2 P_{II}(k)}\,\sqrt{\sum_k P_{TT}(k)}} .
    \label{eq:r_of_W}
\end{equation}
At the optimum, the derivative of $r$ with respect to $W(k)$ vanishes,
\begin{equation}
    \frac{\partial r}{\partial W(k)}
    = \frac{P_{IT}(k)\,\langle L^2\rangle - \langle LT\rangle\,W(k)\,P_{II}(k)}
           {\langle L^2\rangle^{3/2}\,\langle T^2\rangle^{1/2}} = 0 ,
\end{equation}
which yields
\begin{equation}
    W(k) = \frac{\langle L^2\rangle}{\langle LT\rangle}\,
           \frac{P_{IT}(k)}{P_{II}(k)}
         \;\propto\; \frac{P_{IT}(k)}{P_{II}(k)} .
    \label{eq:matched_filter}
\end{equation}
By applying the Cauchy--Schwarz inequality to the numerator of Eq.~(\ref{eq:r_of_W}),
\begin{equation}
    \begin{aligned}
        \sum_k W P_{IT}
        &= \sum_k \bigl(W\sqrt{P_{II}}\bigr)\bigl(P_{IT}/\sqrt{P_{II}}\bigr) \\
        &\le \sqrt{\sum_k W^2 P_{II}}\,\sqrt{\sum_k P_{IT}^2/P_{II}} .
    \end{aligned}
\end{equation}
We find that the optimal filter Eq.~(\ref{eq:matched_filter}) corresponds to the maximum correlation coefficient.
\begin{equation}
    r[W] \le r_{\max} \equiv
    \sqrt{\frac{\sum_k P_{IT}(k)^2/P_{II}(k)}{\sum_k P_{TT}(k)}},
    \label{eq:rmax}
\end{equation}

We first apply the U-net to Gaussian random fields, where the learned mapping, if successful, should converge to an effective matched filter of the form in Eq.~(\ref{eq:matched_filter}) in frequency space, providing a useful sanity check.
When we later apply it to log-normal fields, any improvement beyond the linear estimator can then be interpreted as evidence that the network is using information not captured by the two-point function alone.

\section{Tests on 1D random fields} \label{chap:1d}
In this section, we test the performance of the U-net on 1D toy scenarios, including Gaussian and log-normal random fields.
We first study Gaussian random fields, where an analytic solution is available, as a sanity check.
We then move on to log-normal fields, where no analytic solution is available, but more generally relevant for LSS cross-correlation analyses.
Using these examples, we show how the U-net extracts correlation information and how its behavior can be interpreted in terms of traditional summary statistics.

\subsection{Gaussian random fields with shared seeds}\label{sec:gaussian_shared_seeds}
In the first test case, we choose one Gaussian random field as input to the U-net generated with a given power spectrum $P(k)$, and choose the target field that appears in the loss function to be a low-pass filtered version of the input.
To ensure that the two fields are maximally correlated at low frequencies, we generate them using the same fiducial power spectrum and random seed.
In this case $P_{II} = P$ and $P_{IT} = P_{TT} = P\,\Theta$, where $\Theta$ is the low-pass filter, so $W(k) = \Theta(k)$ and
$\mathbf{L}$ is exactly the low-pass filtered input, i.e. $\mathbf{L}=\mathbf{T}$.
\subsubsection{Data preparation}
We first generate a pair of real-valued Gaussian random fields A1, A2. We start with the same fiducial power spectrum,
\begin{equation}
P(k) = A\frac{\left(|k|/k_0\right)^{n_{\mathrm{low}}}}{1+\left(|k|/k_0\right)^{-n_{\mathrm{high}}+n_{\mathrm{low}}}},
\end{equation}
with $k_0 = 0.01$, $n_{\mathrm{low}} = 1.0$, and $n_{\mathrm{high}} = -1.5$. We then draw complex Fourier coefficients mode by mode as Gaussian random variables with variance fixed by this spectrum,
\begin{equation}
\tilde{\delta}(k)
=
\sqrt{\frac{1}{2}}\,X_k
+
i \sqrt{\frac{1}{2}}\,Y_k,
\qquad
X_k,\,Y_k \sim \mathcal{N}\!\left(0,\sqrt{P(k)}\right),
\end{equation}
so that the real and imaginary parts are independent and the ensemble variance satisfies
\begin{equation}
\bigl\langle |\tilde{\delta}(k)|^2 \bigr\rangle = P(k).
\end{equation}
To ensure that the resulting field in real space is real-valued, the negative-frequency modes are imposed by Hermitian symmetry,
\begin{equation}
\tilde{\delta}(-k)=\tilde{\delta}(k)^{*}.
\end{equation}
The real-space field is then obtained by an inverse fast Fourier transform (IFFT) of these coefficients.
We generate $A_1$ this way, but for $A_2$ we apply a high-frequency cutoff at $k=20$ to the fiducial power spectrum and then repeat the procedure. This ensures that $A_1$ and $A_2$ share the same low-frequency information, while only $A_1$ contains high-frequency modes.
In the first Gaussian case, we let \textbf{I}=$A_1$, \textbf{T}=$A_2$, and choose the same random seed for both. Therefore, the two fields are maximally correlated at low-frequency modes and uncorrelated at high-frequency modes.
We generate 100000 such pairs of 1D Gaussian random fields with 4096 pixels for training, and another 500 pairs for validation, using the U-net architecture described in Chapter \ref{sec:1dunet}. We train the U-net for 10 epochs with a batch size of 32, using the Adam optimizer with a learning rate of $10^{-3}$.
\subsubsection{Results and discussions}
\begin{figure}[!htbp]
	\centering
	\begin{subfigure}{\linewidth}
		\centering
		\includegraphics[width=\linewidth]{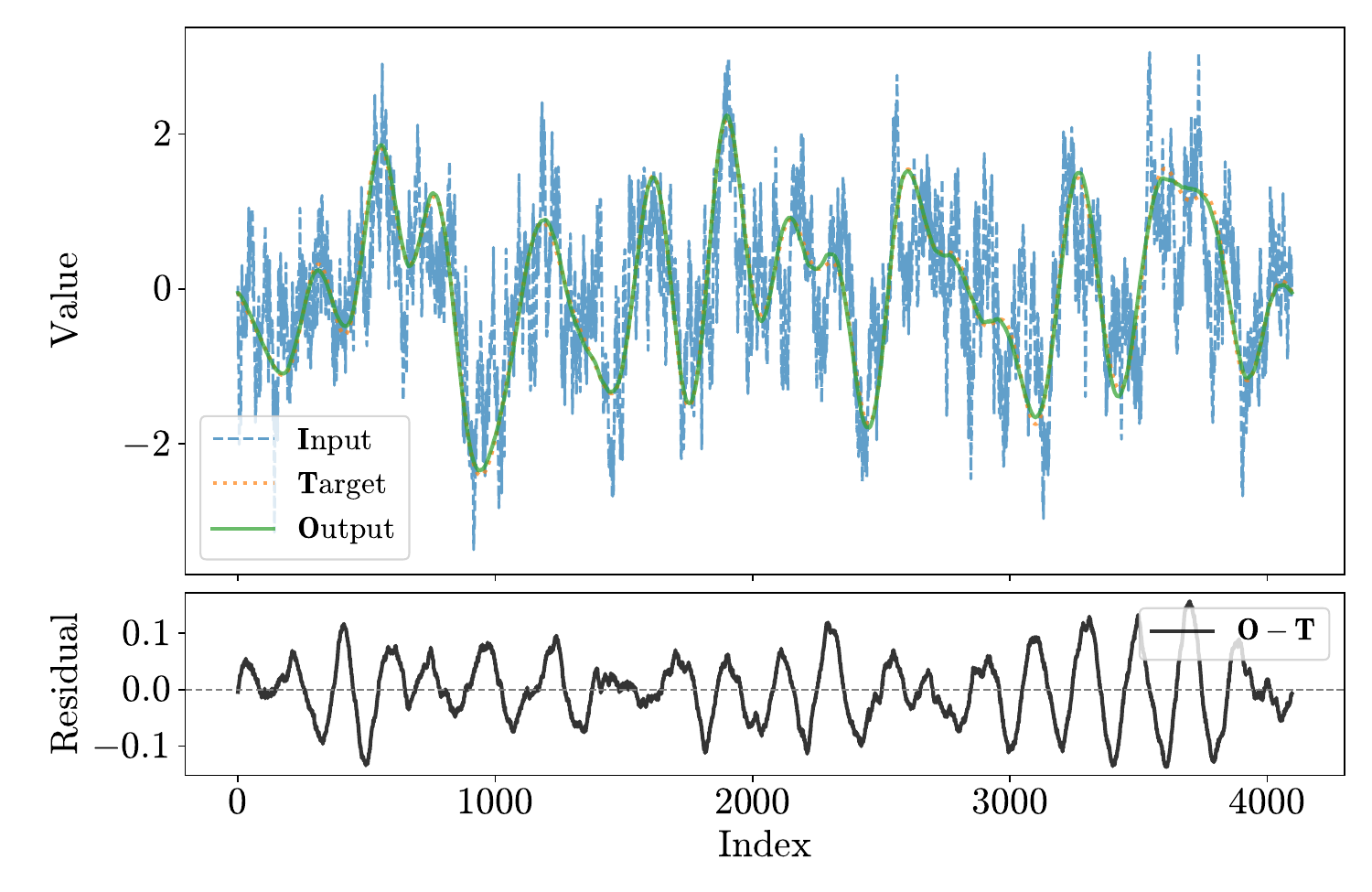}
		\label{fig:gaussian_same_seed_fields}
	\end{subfigure}

	\vspace{0.5em}
	\begin{subfigure}{\linewidth}
		\centering
		\includegraphics[width=\linewidth]{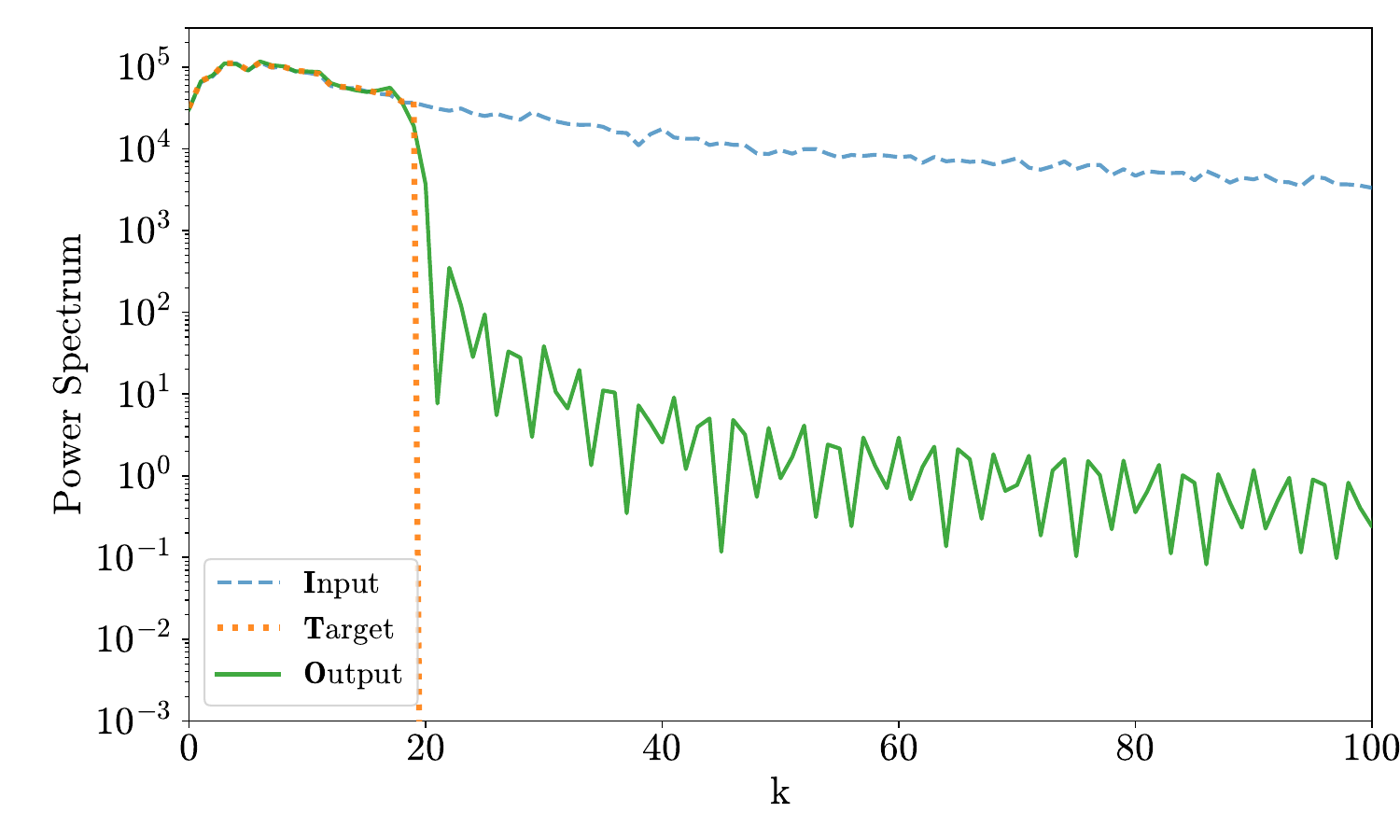}
		\label{fig:gaussian_same_seed_ps_log}
	\end{subfigure}
    \caption{Gaussian-field results for the case $\textbf{I}=A_1$ and $\textbf{T}=A_2$ with the same random seed. Upper panel shows the input, target and output fields from one realization, including the residuals between the output and linear estimator. The oscillatory behavior in the residuals is due to the fact that while the output also has a sharp cutoff in the power spectrum, it is not an exact cutoff, therefore there are residual high-frequency components which causes the oscillatory behavior in the residuals, and frequency of the oscillation is identical to the cutoff scale. Lower panel shows the mean of power spectra averaging from 100 realizations. \textbf{T} and \textbf{O} are normalized to have zero mean and unit variance, while \text{I} are normalized so that it matches the other two at low frequencies. The residuals and cutoff in the power spectra show that the U-net output \textbf{O} is in good agreement with the Gaussian linear estimator \textbf{L}, that is, a low-pass filtered version of the input \textbf{I}.}
	\label{fig:gaussian_same_seed}
\end{figure}

The results are shown in Fig.~\ref{fig:gaussian_same_seed}.
The fields in real space from one realization are displayed in upper panel, while the corresponding power spectra (averaged over 100 realizations) are shown in lower panel. We used a fast Fourier transform (FFT) to compute the power spectra.
In agreement with our expectation, the output \textbf{O} is almost identical to \textbf{T}, which is also evident in the power spectra, where \textbf{O} drops by around 2 orders of magnitude after the cutoff scale.
The oscillatory behavior in the residuals is due to the fact that while the output also has a sharp cutoff in the power spectrum, it is not an exact cutoff, therefore there are residual high-frequency components which causes the oscillatory behavior in the residuals, and frequency of the oscillation is identical to the cutoff scale.
However, this oscillatory behavior is not a concern for the correlation optimization task, as can be seen from the averaged correlation coefficients across 200 realizations shown in Tab.~\ref{tab:gaussian_same_seed_corr}, which shows that the correlation coefficient of \textbf{O} and \textbf{T} is very close to 1.
Note that the correlation coefficient of \textbf{T} and \textbf{L} is not exactly 1, owing to FFT numerical errors and the finite size of the fields, as we apply FFT to \textbf{I}, remove the high-frequency part of the spectrum, and then apply IFFT to obtain \textbf{L}.

\begin{table}[!htbp]
    \centering
    \begin{tabular}{lll}
        \hline
        \textbf{I} vs \textbf{T} & \textbf{L} vs \textbf{T} & \textbf{O} vs \textbf{T}  \\
        \hline
        0.698±0.042 & 0.993±0.006 & 0.992±0.003 \\
        \hline
    \end{tabular}
	\caption{Average Pearson correlation coefficients across 200 realizations, with standard deviation among individual realizations, for the case $\textbf{I}=A_1$ and $\textbf{T}=A_2$ with the same random seed. The quoted uncertainties are scatter expected for a single field, whereas the uncertainty of the mean is smaller by a factor $\sqrt{200}\approx 14$. The same applies to all subsequent tables. The correlation coefficient of output \textbf{O} and target \textbf{T} is very close to 1, indicating that the U-net output is almost identical to the target field, which is a low-pass filtered version of the input. }
	\label{tab:gaussian_same_seed_corr}
\end{table}

\subsection{Gaussian random fields with independent seeds}
\label{sec:gaussian2}
\subsubsection{Data preparation}

In the second Gaussian case, we set \textbf{I}=$A_1+A_2$ and \textbf{T}=$A_2$, where $A_1$ and $A_2$ have different random seeds and are therefore independent Gaussian random fields.
This gives $P_{II} = P\,(1+\Theta)$ while $P_{IT} = P_{TT} = P\,\Theta$, so
$W(k) = \Theta(k)/(1+\Theta(k))$, which equals $1/2$ below the cutoff and
$0$ above, so \textbf{L} is again a low-pass filtered version of \textbf{I}.

\subsubsection{Results and discussions}
\begin{figure}[!htbp]
	\centering
	\begin{subfigure}{\linewidth}
		\centering
		\includegraphics[width=\linewidth]{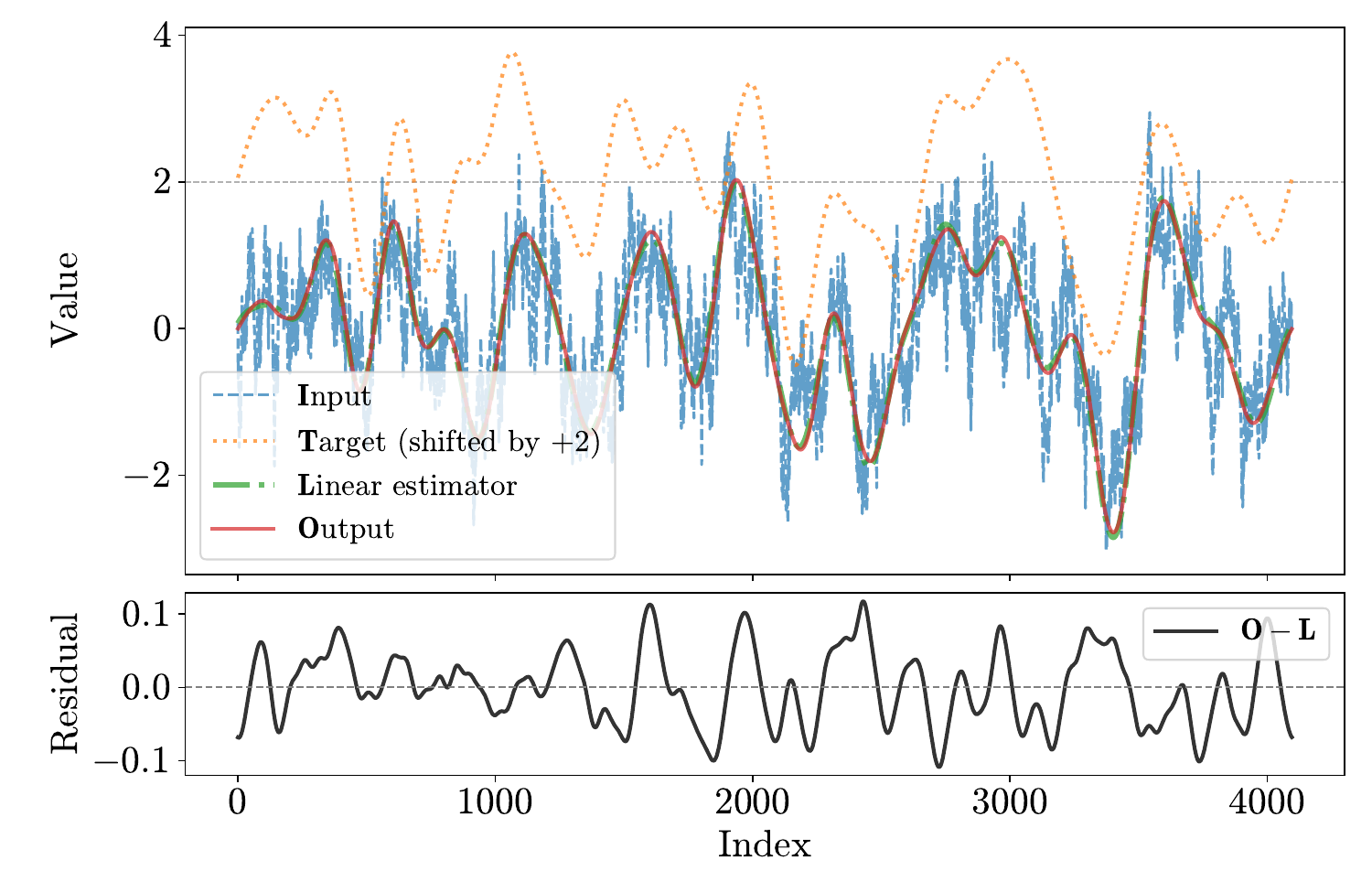}
		\label{fig:gaussian_diff_seed_fields}
	\end{subfigure}

	\vspace{0.5em}
	\begin{subfigure}{\linewidth}
		\centering
		\includegraphics[width=\linewidth]{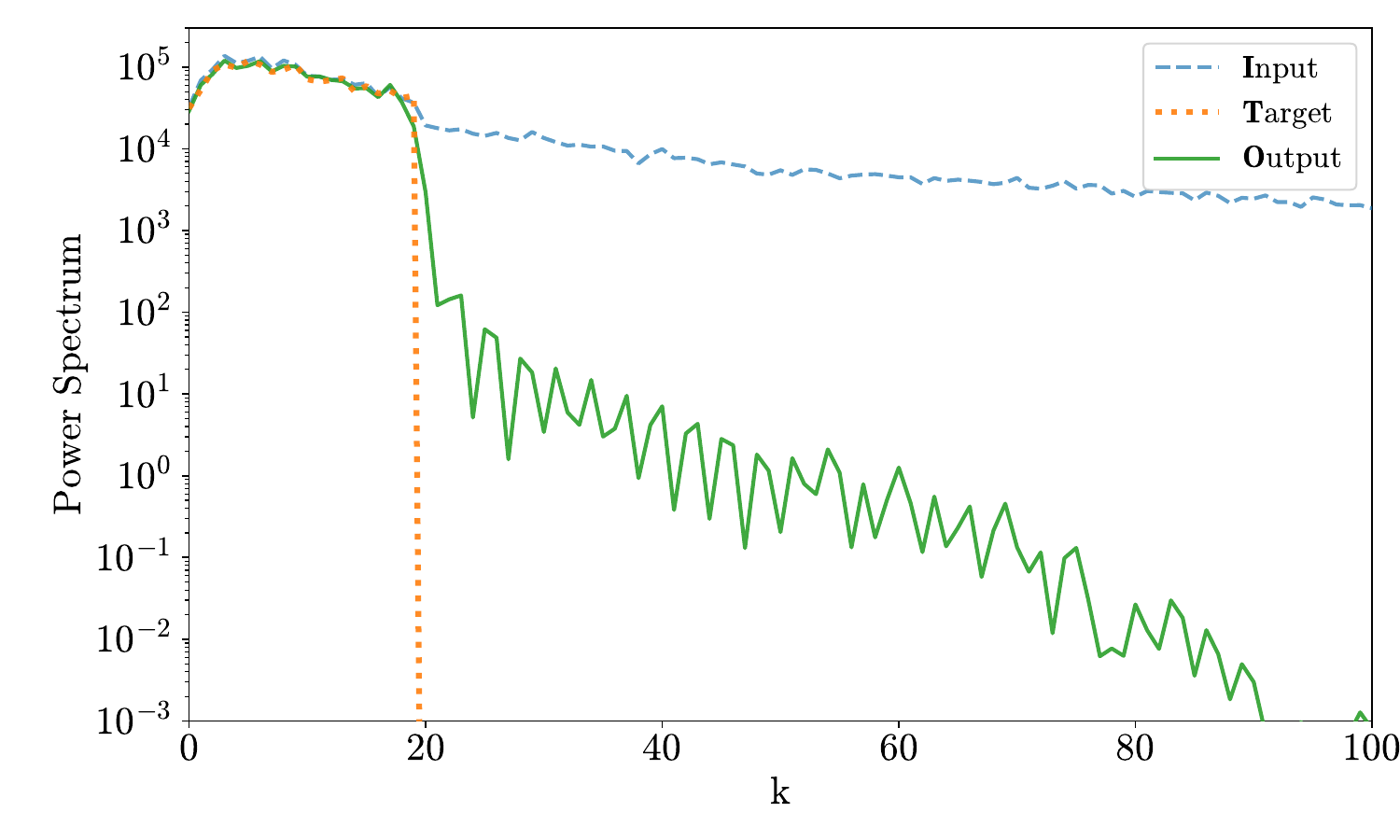}
		\label{fig:gaussian_diff_seed_ps_log}
	\end{subfigure}
    \caption{Gaussian-field results for the case $\textbf{I}=A_1+A_2$ and $\textbf{T}=A_2$ with independent random seeds. Upper panel shows the input, target, output and linear estimator fields from one realization, including the residuals between the output and linear estimator. Lower panel shows the mean of power spectra averaging from 100 realizations. \textbf{T} is normalized to have zero mean and unit variance, while \textbf{I} and \textbf{O} are normalized so that they match \textbf{T} at low frequencies. The residuals and cutoff in the power spectra show that the U-net output \textbf{O} is in good agreement with the Gaussian linear estimator \textbf{L}, that is, a low-pass filtered version of the input \textbf{I}.}
	\label{fig:gaussian_diff_seed}
\end{figure}

The results are shown in Fig.~\ref{fig:gaussian_diff_seed}. The fields in real space from one realization are displayed in upper panel, while the corresponding power spectra (averaged over 100 realizations) are shown in lower panel.
Again we also find that the output \textbf{O} has a drastic decrease in power spectrum after the cutoff point, which is consistent with our expectation. The averaged correlation coefficients across 200 realizations are shown in Tab.~\ref{tab:gaussian_diff_seed_corr}, which is also in agreement with the above conclusion.

\begin{table}[!htbp]
    \centering
    \begin{tabular}{lll}
        \hline
        \textbf{I} vs \textbf{T} & \textbf{L} vs \textbf{T} & \textbf{O} vs \textbf{T}  \\
        \hline
        0.557±0.091 & 0.688±0.094 & 0.686±0.094 \\
        \hline
    \end{tabular}
	\caption{Average Pearson correlation coefficients across 200 realizations, with standard deviation among different realizations, for the case $\textbf{I}=A_1+A_2$ and $\textbf{T}=A_2$ with independent random seeds.}
	\label{tab:gaussian_diff_seed_corr}
\end{table}

These two cases confirm that the U-net is able to extract the correlation structure between two fields, which in these Gaussian examples can be interpreted as a matched filter in Fourier space.
This is not trivial: \citet{Betzler_2020} found that simple convolutional networks, even when extended with fully connected layers, failed to detect a single-frequency signal in Gaussian noise from position-space data, although the same task is easy in Fourier space.
While their classification task differs from our correlation-optimization setup, the large convolution kernels and multi-scale structure of the U-net, together with a loss that directly rewards correlation with the target, appear to make the optimal frequency-space filter accessible from real-space input alone.

\subsection{Log-normal fields as input and target}
\label{sec:log1}
While the CMB anisotropies are highly Gaussian, the large-scale structure of the Universe is non-Gaussian due to nonlinear gravitational evolution.
This scenario leads to cosmological cross-correlation analyses of interest, where the highly Gaussian CMB maps are cross-correlated with non-Gaussian LSS tracer maps, such as galaxy density fields or weak lensing convergence fields. In such cases, gains over the linear matched filter approach are in principle possible.

The LSS tracer fields can be approximated as log-normal fields, which has long been used as an effective approximation to mildly nonlinear matter and galaxy density fields \citep[e.g.][]{Coles_Jones_1991,Friedrich2018}.
We therefore further test the performance of the U-net on log-normal fields, for which no simple analytic solutions are available, in order to gain further insight into how the U-net performs on more realistic LSS tracer maps.

\subsubsection{Data preparation}

As before, we start with a pair of 1D Gaussian random fields $A_1$ and $A_2$ with the same fiducial power spectrum, except for a high-frequency cutoff applied to the power spectrum of $A_2$, and we choose them to have different random seeds.
In the first log-normal case, we choose the input field to be \textbf{I}=$100A_1+\exp(50A_2)$ and the target field to be \textbf{T}=$\exp(50A_2)$.
The coefficients are arbitrarily chosen, and the results show that this choice ensures that the non-Gaussianity is sufficiently strong.
Linear estimator \textbf{L} is constructed by Fourier transforming the input field \textbf{I}, multiply by the ratio of the cross spectrum of \textbf{I} and \textbf{T} over the auto spectrum of \textbf{I}, and then inverse fourier transforming back to real space. The resulting field deviates slightly from low-pass filtered version of \textbf{I}, but the difference is very small, because the leakage to high frequency induced by the log-normal transformation is relatively small.
In contrast to the Gaussian case, although A1 and A2 are independent (uncorrelated), the log-normal transformation introduces non-Gaussian structure and mode coupling in the target field.
As a result, the mixed input field \textbf{I} contains information that is not optimally extracted by a linear estimator \textbf{L}, leaving room for the U-net to learn non-Gaussian information that improves the correlation with the target \textbf{T}.

We generate 100000 such pairs of 1D log-normal fields with 4096 pixels for training, and another 500 pairs for validation. We train the U-net for 10 epochs, using the same architecture.

\subsubsection{Results and discussions}
\label{sec:para}

\begin{figure}[!htbp]
	\centering
	\begin{subfigure}{\linewidth}
		\centering
		\includegraphics[width=\linewidth]{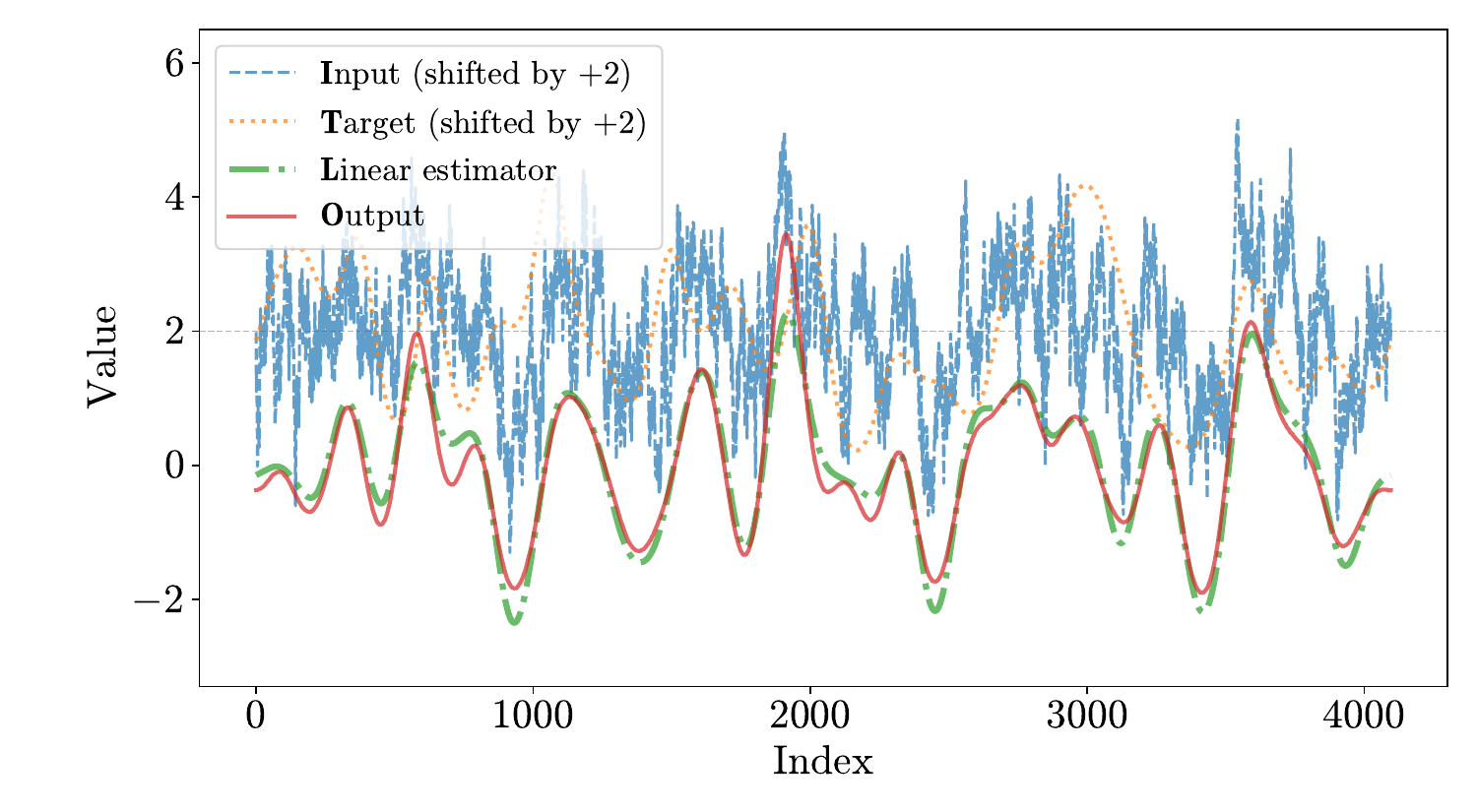}
		\label{fig:log1_fields}
	\end{subfigure}

	\vspace{0.5em}
	\begin{subfigure}{\linewidth}
		\centering
		\includegraphics[width=\linewidth]{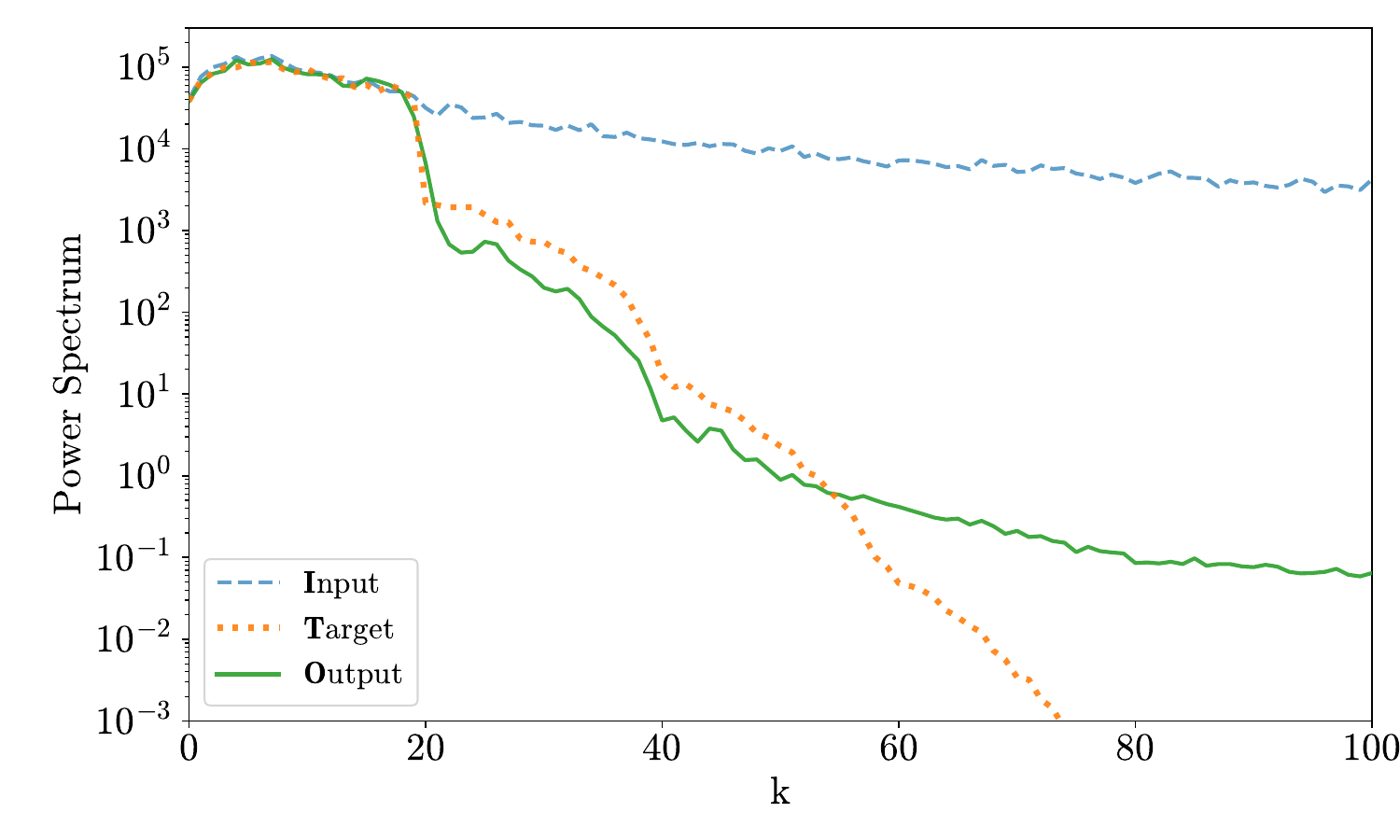}
		\label{fig:log1_ps}
	\end{subfigure}
    \caption{Log-normal-field results for the case $\textbf{I}=100A_1+\exp(50A_2)$ and $\textbf{T}=\exp(50A_2)$. Upper panel shows the input, target, output and linear estimator fields from one realization. Lower panel shows the mean of power spectra averaging from 100 realizations. The fields are normalized to have zero mean and unit variance. The residuals and cutoff in the power spectra show that the U-net output \textbf{O} is in good agreement with the Gaussian linear estimator \textbf{L}, that is, a low-pass filtered version of the input \textbf{I}.}
	\label{fig:log1}
\end{figure}

We repeat the analysis for the log-normal case with \textbf{I}=$100A_1+\exp(50A_2)$ and \textbf{T}=$\exp(50A_2)$. The upper panel of Fig.~\ref{fig:log1} shows the input \textbf{I}, target field \textbf{T}, linear estimator (the analytic solution assuming Gaussianity) \textbf{L}, and U-net output \textbf{O} in real space from one realization, while the corresponding power spectra (averaged over 100 realizations) are shown in lower panel.
We see that although the output \textbf{O} also exhibits a sharp decrease in power spectrum, it deviates from the linear estimator \textbf{L}.
If both fields were Gaussian, we would expect the analytic solution to be the linear estimator \textbf{L}, which is a low-pass filtered version of the input as in the previous case. Since the fields here are log-normal, departures from the linear estimator suggest that the U-net is also sensitive to nonlinear information beyond the power spectrum alone.
We also compare the average correlation coefficients across 200 realizations in Tab.~\ref{tab:log1_corr}. The U-net output \textbf{O} has a substantially higher correlation with the target field \textbf{T} than the linear estimator \textbf{L}, indicating that the network is using non-linear information.

\begin{figure}[!htbp]
	\centering
	\includegraphics[width=\linewidth]{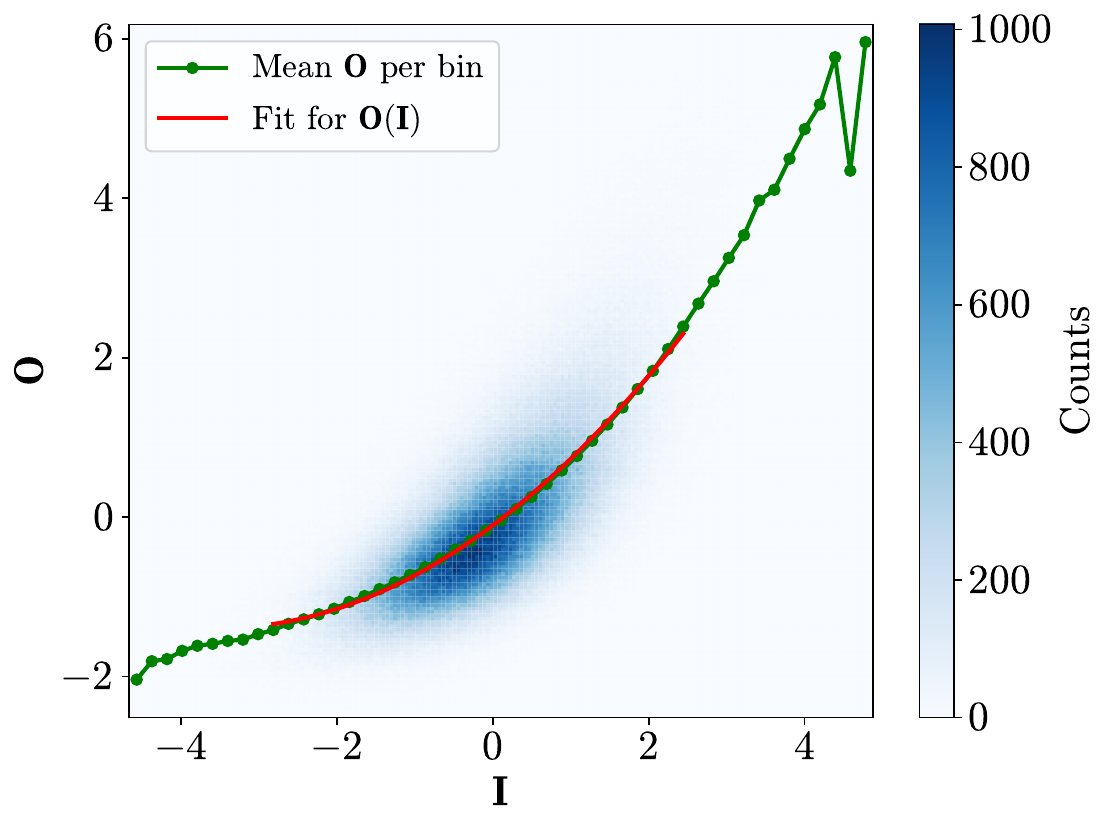}
	\caption{The 2D histogram of normalized input \textbf{I} and output \textbf{O}, stacking 400 realizations, for the case $\textbf{I}=100A_1+\exp(50A_2)$ and $\textbf{T}=\exp(50A_2)$. After taking the mean value of \textbf{O} with each \textbf{I} bin, the result can be well fitted with a parabolic function $\textbf{O} = 0.11 \textbf{I}^2 + 0.71 \textbf{I} - 0.11$.}
	\label{fig:parabola}
\end{figure}

\begin{table}[!htbp]
    \centering
    \begin{tabular}{llll}
        \hline
        \textbf{I} vs \textbf{T} & \textbf{L} vs \textbf{T} & \textbf{O} vs \textbf{T}  & \textbf{PF} vs \textbf{T}  \\
        \hline
        0.336±0.116 & 0.452±0.138 & 0.465±0.152 & 0.460±0.145 \\
        \hline
    \end{tabular}
	\caption{Average Pearson correlation coefficients across 200 realizations, with standard deviation among different realizations, for the case $\textbf{I}=100A_1+\exp(50A_2)$ and $\textbf{T}=\exp(50A_2)$. $\textbf{PF}$ denotes the parabolic transformed and then low-pass filtered version of the input \textbf{I}.}
	\label{tab:log1_corr}
\end{table}
Another interesting observation is that when we plot the 2D histogram of the input \textbf{I} and output \textbf{O} (normalized to have zero mean and unit variance) in Fig.~\ref{fig:parabola}, we find that the distribution can be well fitted by a parabolic function:
\begin{equation}
	\textbf{O} = 0.11 \textbf{I}^2 + 0.71 \textbf{I} - 0.11
	\label{eq:parabola_fit}
\end{equation}

This quadratic response is the lowest-order local nonlinear correction and is qualitatively reminiscent of statistics sensitive to mode coupling, such as integrated or squeezed-bispectrum-type information \citep{Chiang_2015, Halder2024}.
\begin{figure}[!htbp]
	\centering
	\includegraphics[width=\linewidth]{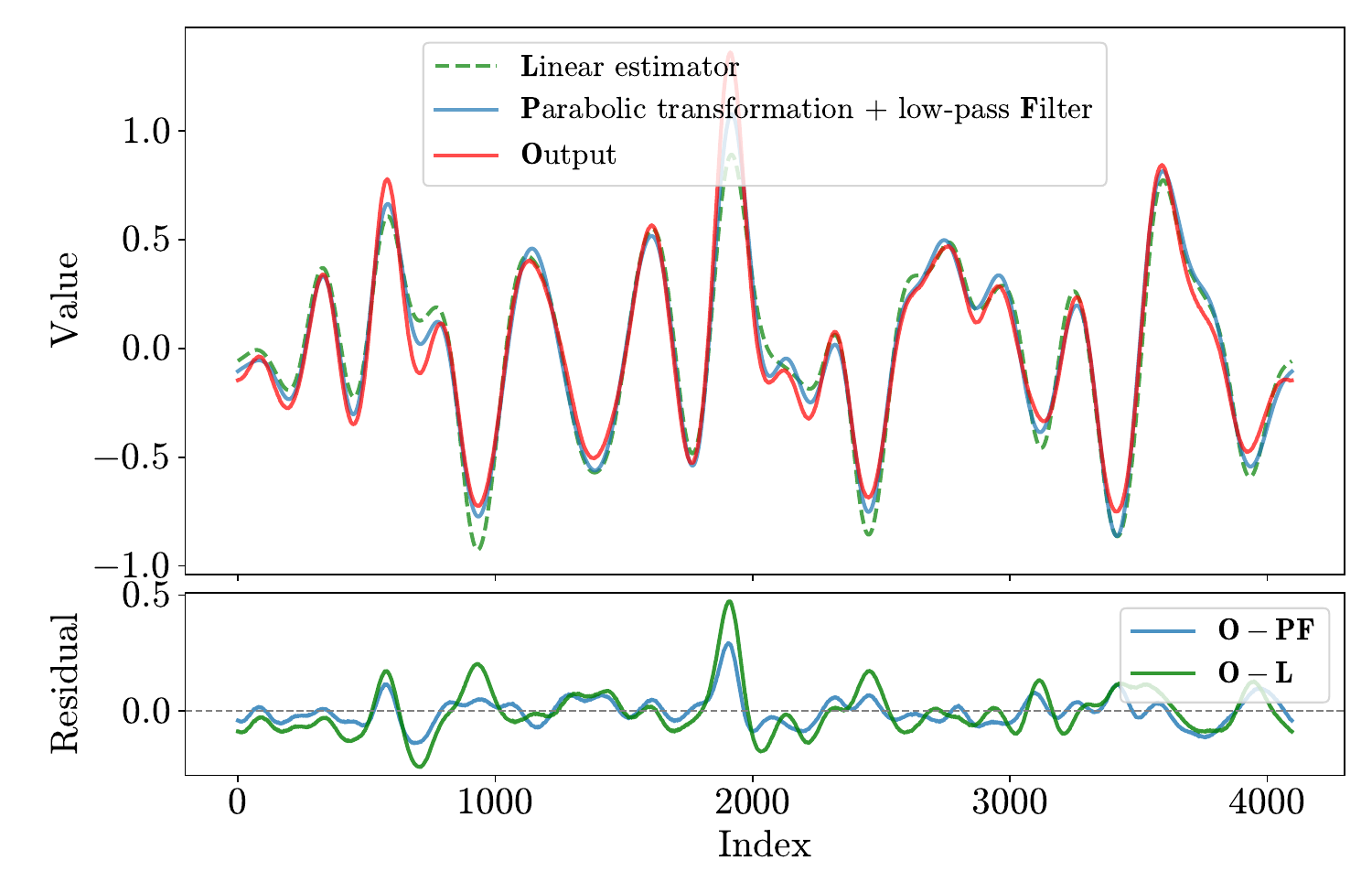}
	\caption{Fields in real space with residuals for the case $\textbf{I}=100A_1+\exp(50A_2)$ and $\textbf{T}=\exp(50A_2)$. \textbf{I} is the input, \textbf{T} is the target field, \textbf{O} is the U-net output, and \textbf{PF} denotes a parabolic transformation of \textbf{I} followed by a low-pass filter.}
	\label{fig:log1_fields5}
\end{figure}
Therefore, a reasonable interpretation is that, to leading order, the U-net output can be approximated as a combination of a frequency filter, which captures Gaussian two-point information, and a quadratic local transform, which captures part of the nonlinear response.

To validate this hypothesis, we attempted to recreate the U-net output using a combination of parabolic transformation and linear filtering.
We define \textbf{PF} as the result of applying the parabolic transformation in Eq.~\ref{eq:parabola_fit} to the input \textbf{I} and then applying low-pass filtering.
We plot the fields in Fig.~\ref{fig:log1_fields5} and compute the average correlation coefficients with the target field \textbf{T} in Tab.~\ref{tab:log1_corr}.
The results show that \textbf{PF} provides good approximations to the U-net output \textbf{O}, supporting the heuristic interpretation that the network combines power-spectrum analysis (linear filtering) with integrated-bispectrum-like analysis (parabolic transformation).

\subsection{Log-normal fields as input and Gaussian fields as target}
\label{sec:log2}
\subsubsection{Data preparation}
In the second log-normal case, we choose the input to be log-normal while the target remains Gaussian, which is analogous to the cross-correlation case of a galaxy overdensity field and an ISW field: \textbf{I}=$\exp(50A_1)+100A_2$, \textbf{T}=$A_2$.
As in the previous case, the log-normal transformation induces non-Gaussian structure and mode coupling, so we expect that the U-net may use information beyond the power spectrum to improve the correlation with the target \textbf{T}.
\subsubsection{Results and discussions}
\begin{figure}[!htbp]
	\centering
	\begin{subfigure}{\linewidth}
		\centering
		\includegraphics[width=\linewidth]{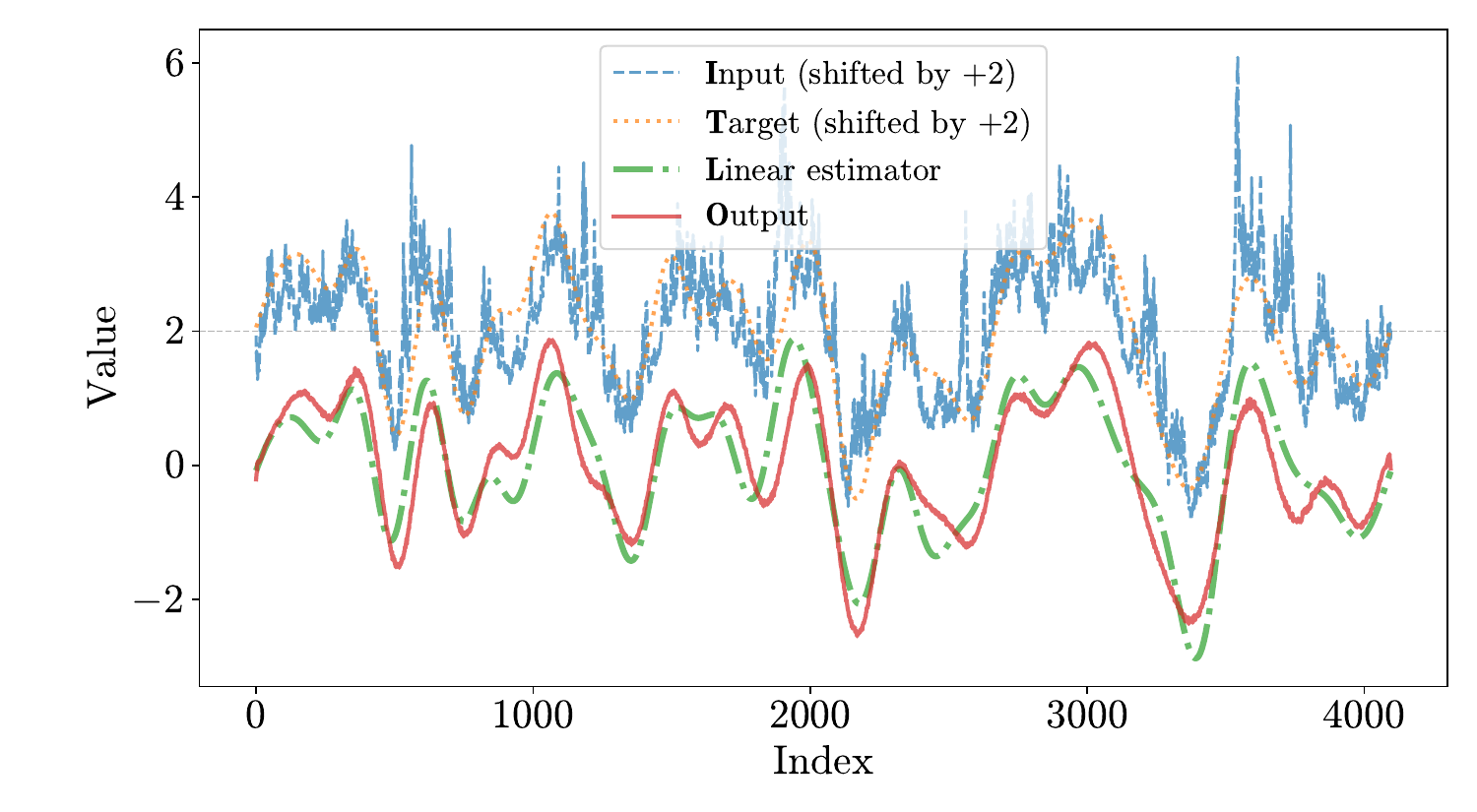}
		\label{fig:log2_fields}
	\end{subfigure}

	\vspace{0.5em}
	\begin{subfigure}{\linewidth}
		\centering
		\includegraphics[width=\linewidth]{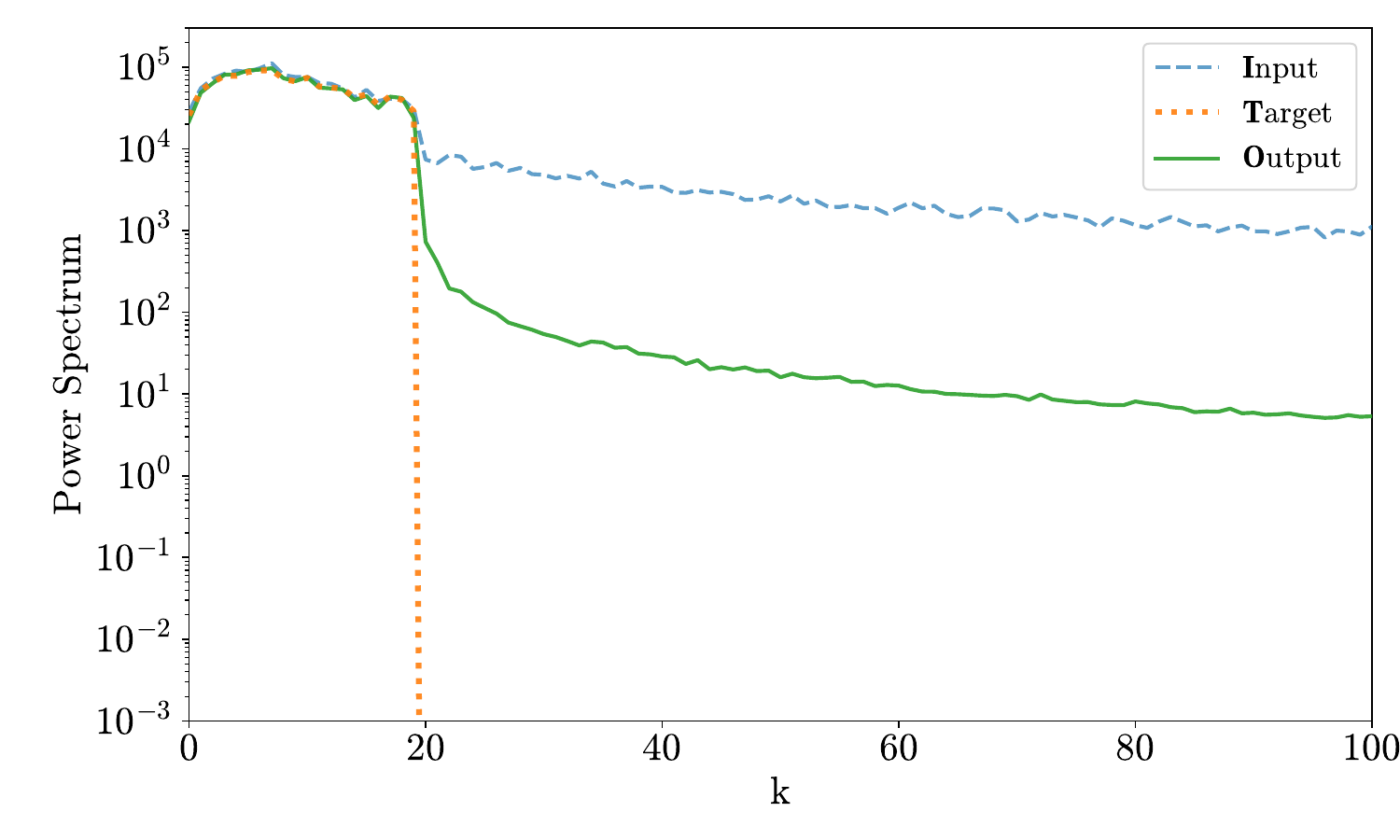}
		\label{fig:log2_ps_log}
	\end{subfigure}
    \caption{Log-normal-field results for the case $\textbf{I}=\exp(50A_1)+100A_2$ and $\textbf{T}=A_2$. Upper panel shows the input, target, output and linear estimator fields from one realization, including the residuals between the output and linear estimator. The oscillatory behavior in the residuals is due to the fact that while the output also has a sharp cutoff in the power spectrum, it is not an exact cutoff, therefore there are residual high-frequency components which causes the oscillatory behavior in the residuals, and frequency of the oscillation is identical to the cutoff scale. Lower panel shows the mean of power spectra averaging from 100 realizations. The fields are normalized to have zero mean and unit variance. The residuals and cutoff in the power spectra show that the U-net output \textbf{O} is in good agreement with the Gaussian linear estimator \textbf{L}, that is, a low-pass filtered version of the input \textbf{I}.}
	\label{fig:log2}
\end{figure}

The fields in real space from one realization and the power spectra (averaged over 100 realizations) are shown in the upper and lower panels of Fig.~\ref{fig:log2}, respectively, while the average correlation coefficients across 200 realizations are shown in Tab.~\ref{tab:log2_corr}.

\begin{table}[!htbp]
    \centering
    \begin{tabular}{lll}
        \hline
        \textbf{I} vs \textbf{T} & \textbf{L} vs \textbf{T} & \textbf{O} vs \textbf{T} \\
        \hline
        0.755±0.059 & 0.857±0.052 & 0.953±0.015\\
        \hline
    \end{tabular}
	\caption{Average Pearson correlation coefficients across 200 realizations, with standard deviation among different realizations, for the case $\textbf{I}=\exp(50A_1)+100A_2$ and $\textbf{T}=A_2$.}
	\label{tab:log2_corr}
\end{table}
In this case the 2D histogram of the input \textbf{I} and output \textbf{O} does not exhibit a parabolic feature.
However, we still find that the U-net output \textbf{O} has a much higher correlation with the target field \textbf{T} (0.953) than the linear estimator \textbf{L} (0.857), indicating that the network is using non-Gaussian information not captured by the linear Gaussian estimator alone.

\section{Conclusions} \label{chap:conclusion}
This work is motivated by the insufficiency of 2-point statistics in cosmological cross-correlation analyses of non-Gaussian random fields.
We proposed a data-driven method that uses a U-net to learn a correlation-optimized tracer field from an input tracer field.
We tested the method on 1D toy models where the optimal solution is analytically known.

Our main findings are as follows. (i) For Gaussian random fields, the U-net converges to a matched filter in frequency space, reproducing the analytic linear estimator (Sect. \ref{sec:gaussian_shared_seeds}, \ref{sec:gaussian2}), which validates the setup.
(ii) For a log-normal input and a Gaussian target, which is relevant for understanding how it may perform on realistic cross-correlation analyses involving the non-Gaussian matter density field,
the U-net raises the mean Pearson correlation coefficient with the target from $0.857$ to $0.953$ compared to a linear estimator (Sect.~\ref{sec:log2}),
demonstrating that the U-net can exploit information beyond the power spectrum.
(iii) For a log-normal target, the output is well approximated by a combination of linear filtering and a local quadratic transformation (Sect. \ref{sec:para}),
suggesting that the correlation-optimized tracer combines power-spectrum information with a
mode-coupling response reminiscent of squeezed-bispectrum statistics.

The results show that the U-net can increase the correlation signal by extracting information beyond the power spectrum, potentially improving the SNR in cosmological cross-correlation analyses.
The framework can be subsequently applied to more realistic 2D and spherical fields, such as galaxy overdensity maps, ISW maps and CMB lensing maps, to quantify and gain in detection significance in ISW and CMB lensing cross-correlation analyses \citep{Planck_2016_ISW, Farren_2024}.

\section*{Data availability}
The code used to generate the random fields, train the U-net, and reproduce the results presented in this paper is publicly available at \url{https://github.com/Franz352/unet_paper}.

\begin{acknowledgements}
We thank Nikolas Frediani for helpful discussions.
Funded by the Deutsche Forschungsgemeinschaft (DFG, German Research Foundation) under Germany's Excellence Strategy – EXC-2094/2 – 390783311.

The Large-Scale Structure (LSS) research group at Konkoly Observatory has been supported by a \emph{Lend\"ulet} excellence grant by the Hungarian Academy of Sciences (MTA). This project has received funding from the European Union’s Horizon Europe research and innovation programme under the Marie Skłodowska-Curie grant agreement number 101130774. Funding for this project was also available in part through the Hungarian National Research, Development and Innovation Office (NKFIH, grant OTKA NN147550).
\end{acknowledgements}

\bibliographystyle{bibtex/aa}
\bibliography{refs_clean}

\begin{thebibliography}{41}
\expandafter\ifx\csname natexlab\endcsname\relax\def\natexlab#1{#1}\fi

\bibitem[{Alam {et~al.}(2021)}]{Alam_2021}
Alam, S. {et~al.} 2021,
  \href{https://doi.org/10.1103/PhysRevD.103.083533}{Phys. Rev. D, 103, 083533}

\bibitem[{Amon {et~al.}(2022)}]{Amon_2022}
Amon, A. {et~al.} 2022,
  \href{https://doi.org/10.1103/PhysRevD.105.023514}{Phys. Rev. D, 105, 023514}

\bibitem[{Betzler \& Krippendorf(2020)}]{Betzler_2020}
Betzler, P. \& Krippendorf, S. 2020,
  \href{https://doi.org/10.1002/prop.202000022}{Fortschritte der Physik, 68,
  2000022}

\bibitem[{Brieden {et~al.}(2023)Brieden, Gil-Mar{\'i}n, \&
  Verde}]{Brieden_2023}
Brieden, S., Gil-Mar{\'i}n, H., \& Verde, L. 2023,
  \href{https://doi.org/10.1088/1475-7516/2023/04/023}{JCAP, 2023, 023}

\bibitem[{Carron {et~al.}(2022)Carron, Mirmelstein, \& Lewis}]{Carron_2022}
Carron, J., Mirmelstein, M., \& Lewis, A. 2022,
  \href{https://doi.org/10.1088/1475-7516/2022/09/039}{JCAP, 2022, 039}

\bibitem[{Carron \& Szapudi(2013)}]{Carron_2013}
Carron, J. \& Szapudi, I. 2013,
  \href{http://dx.doi.org/10.1093/mnrasl/slt167}{Monthly Notices of the Royal
  Astronomical Society: Letters, 439, L11}

\bibitem[{{Chang} {et~al.}(2025){Chang}, {Nayak}, {Walther}, \&
  {Gruen}}]{Chang_2025}
{Chang}, S., {Nayak}, P., {Walther}, M., \& {Gruen}, D. 2025,
  \href{https://doi.org/10.48550/arXiv.2508.03264}{arXiv e-prints,
  arXiv:2508.03264}

\bibitem[{Chiang {et~al.}(2015)Chiang, Wagner, S{\'a}nchez, Schmidt, \&
  Komatsu}]{Chiang_2015}
Chiang, C.-T., Wagner, C., S{\'a}nchez, A.~G., Schmidt, F., \& Komatsu, E.
  2015, \href{https://doi.org/10.1088/1475-7516/2015/09/028}{JCAP, 2015, 028}

\bibitem[{Coles \& Jones(1991)}]{Coles_Jones_1991}
Coles, P. \& Jones, B. 1991,
  \href{https://doi.org/10.1093/mnras/248.1.1}{MNRAS, 248, 1}

\bibitem[{Cowell {et~al.}(2024)Cowell, Alonso, \& Liu}]{Cowell_2024}
Cowell, J.~A., Alonso, D., \& Liu, J. 2024,
  \href{https://doi.org/10.1093/mnras/stae2492}{MNRAS, 535, 3129}

\bibitem[{Cowell {et~al.}(2026)Cowell, Armijo, Thiele, Marques, Novaes,
  Grand{\'o}n, Cheng, Shirasaki, Alonso, \& Liu}]{Cowell_2025}
Cowell, J.~A., Armijo, J., Thiele, L., {et~al.} 2026,
  \href{https://doi.org/10.1093/mnras/stag033}{MNRAS, 546, stag033}

\bibitem[{{DESI Collaboration} {et~al.}(2025{\natexlab{a}})}]{DESI_2025_FS}
{DESI Collaboration} {et~al.} 2025{\natexlab{a}},
  \href{https://doi.org/10.1088/1475-7516/2025/07/028}{JCAP, 2025, 028}

\bibitem[{{DESI Collaboration} {et~al.}(2025{\natexlab{b}})}]{DESI_2025}
{DESI Collaboration} {et~al.} 2025{\natexlab{b}},
  \href{https://doi.org/10.1103/tr6y-kpc6}{Phys. Rev. D, 112, 083515}

\bibitem[{Dong {et~al.}(2022)Dong, Zhang, Sun, \& Park}]{Dong_2022}
Dong, F., Zhang, P., Sun, Z., \& Park, C. 2022,
  \href{https://doi.org/10.3847/1538-4357/ac905b}{ApJ, 938, 72}

\bibitem[{Ebina \& White(2025)}]{EbinaWhite2025}
Ebina, H. \& White, M. 2025,
  \href{https://doi.org/10.1088/1475-7516/2025/01/150}{JCAP, 2025, 150}

\bibitem[{Farren {et~al.}(2024)Farren, Krolewski, MacCrann,
  {et~al.}}]{Farren_2024}
Farren, G.~S., Krolewski, A., MacCrann, N., {et~al.} 2024,
  \href{https://doi.org/10.3847/1538-4357/ad31a5}{ApJ, 966, 157}

\bibitem[{Fluri {et~al.}(2018)Fluri, Kacprzak, Refregier, Amara, Lucchi, \&
  Hofmann}]{Fluri_2018}
Fluri, J., Kacprzak, T., Refregier, A., {et~al.} 2018,
  \href{https://doi.org/10.1103/PhysRevD.98.123518}{Phys. Rev. D, 98, 123518}

\bibitem[{{Friedrich} {et~al.}(2018){Friedrich}, {Gruen}, {DeRose}, {Kirk},
  {Krause}, {McClintock}, {Rykoff}, {Seitz}, {Wechsler}, {Bernstein}, {Blazek},
  {Chang}, {Hilbert}, {Jain}, {Kovacs}, {Lahav}, {Abdalla}, {Allam}, {Annis},
  {Bechtol}, {Benoit-L{\'e}vy}, {Bertin}, {Brooks}, {Carnero Rosell}, {Carrasco
  Kind}, {Carretero}, {Cunha}, {D'Andrea}, {da Costa}, {Davis}, {Desai},
  {Diehl}, {Dietrich}, {Drlica-Wagner}, {Eifler}, {Fosalba}, {Frieman},
  {Garc{\'\i}a-Bellido}, {Gaztanaga}, {Gerdes}, {Giannantonio}, {Gruendl},
  {Gschwend}, {Gutierrez}, {Honscheid}, {James}, {Jarvis}, {Kuehn},
  {Kuropatkin}, {Lima}, {March}, {Marshall}, {Melchior}, {Menanteau}, {Miquel},
  {Mohr}, {Nord}, {Plazas}, {Sanchez}, {Scarpine}, {Schindler}, {Schubnell},
  {Sevilla-Noarbe}, {Sheldon}, {Smith}, {Soares-Santos}, {Sobreira}, {Suchyta},
  {Swanson}, {Tarle}, {Thomas}, {Troxel}, {Vikram}, {Weller}, \& {DES
  Collaboration}}]{Friedrich2018}
{Friedrich}, O., {Gruen}, D., {DeRose}, J., {et~al.} 2018,
  \href{https://doi.org/10.1103/PhysRevD.98.023508}{\prd, 98, 023508}

\bibitem[{Gao {et~al.}(2023)Gao, Li, Ni, \& Zhang}]{Gao_2023}
Gao, L.-Y., Li, Y., Ni, S., \& Zhang, X. 2023,
  \href{https://doi.org/10.1093/mnras/stad2646}{MNRAS, 525, 5278}

\bibitem[{Ge {et~al.}(2025)}]{Ge_2025}
Ge, F. {et~al.} 2025, \href{https://doi.org/10.1103/PhysRevD.111.083534}{Phys.
  Rev. D, 111, 083534}

\bibitem[{Gillet {et~al.}(2019)Gillet, Mesinger, Greig, Liu, \&
  Ucci}]{Gillet_2019}
Gillet, N., Mesinger, A., Greig, B., Liu, A., \& Ucci, G. 2019,
  \href{https://doi.org/10.1093/mnras/stz010}{MNRAS, 484, 282}

\bibitem[{Gong {et~al.}(2024)Gong, Halder, Bohrdt, Seitz, \&
  Gebauer}]{Gong_2024}
Gong, Z., Halder, A., Bohrdt, A., Seitz, S., \& Gebauer, D. 2024,
  \href{https://doi.org/10.3847/1538-4357/ad582e}{ApJ, 971, 156}

\bibitem[{Gupta {et~al.}(2018)Gupta, Zorrilla~Matilla, Hsu, \&
  Haiman}]{Gupta_2018}
Gupta, A., Zorrilla~Matilla, J.~M., Hsu, D., \& Haiman, Z. 2018,
  \href{https://doi.org/10.1103/PhysRevD.97.103515}{Phys. Rev. D, 97, 103515}

\bibitem[{Halder(2024)}]{Halder2024}
Halder, A. 2024, PhD thesis, Munich U.,
  \href{https://doi.org/10.5282/edoc.33227}{doi:10.5282/edoc.33227}

\bibitem[{Jeffrey {et~al.}(2020)Jeffrey, Lanusse, Lahav, \&
  Starck}]{Jeffrey_2020}
Jeffrey, N., Lanusse, F., Lahav, O., \& Starck, J.-L. 2020,
  \href{https://doi.org/10.1093/mnras/staa127}{MNRAS, 492, 5023}

\bibitem[{Jeffrey {et~al.}(2025)Jeffrey, Whiteway, Gatti, Williamson, Alsing,
  Porqueres, {et~al.}}]{Jeffrey_2025}
Jeffrey, N., Whiteway, L., Gatti, M., {et~al.} 2025,
  \href{https://doi.org/10.1093/mnras/stae2629}{MNRAS, 536, 1303}

\bibitem[{Kacprzak \& Fluri(2022)}]{Kacprzak_2022}
Kacprzak, T. \& Fluri, J. 2022,
  \href{https://doi.org/10.1103/PhysRevX.12.031029}{Physical Review X, 12,
  031029}

\bibitem[{Krolewski {et~al.}(2021)Krolewski, Ferraro, \&
  White}]{Krolewski_2021}
Krolewski, A., Ferraro, S., \& White, M. 2021,
  \href{https://doi.org/10.1088/1475-7516/2021/12/028}{JCAP, 2021, 028}

\bibitem[{LeCun {et~al.}(2015)LeCun, Bengio, \& Hinton}]{LeCun_2015}
LeCun, Y., Bengio, Y., \& Hinton, G. 2015,
  \href{https://doi.org/10.1038/nature14539}{Nature, 521, 436}

\bibitem[{Li {et~al.}(2023)}]{Li_2023}
Li, X. {et~al.} 2023, \href{https://doi.org/10.1103/PhysRevD.108.123518}{Phys.
  Rev. D, 108, 123518}

\bibitem[{Lu {et~al.}(2023)Lu, Haiman, \& Li}]{Lu_2023}
Lu, T., Haiman, Z., \& Li, X. 2023,
  \href{https://doi.org/10.1093/mnras/stad686}{MNRAS, 521, 2050}

\bibitem[{Madhavacheril {et~al.}(2024)}]{Madhavacheril_2024}
Madhavacheril, M.~S. {et~al.} 2024,
  \href{https://doi.org/10.3847/1538-4357/acff5f}{ApJ, 962, 113}

\bibitem[{Nayak {et~al.}(2024)Nayak, Walther, Gruen, \& Adiraju}]{Nayak_2024}
Nayak, P., Walther, M., Gruen, D., \& Adiraju, S. 2024,
  \href{https://doi.org/10.1051/0004-6361/202348485}{A\&A, 689, A153}

\bibitem[{Neyrinck {et~al.}(2009)Neyrinck, Szapudi, \& Szalay}]{Neyrinck_2009}
Neyrinck, M.~C., Szapudi, I., \& Szalay, A.~S. 2009,
  \href{http://dx.doi.org/10.1088/0004-637X/698/2/L90}{ApJ, 698, L90}

\bibitem[{Perreault~Levasseur {et~al.}(2017)Perreault~Levasseur, Hezaveh, \&
  Wechsler}]{PerreaultLevasseur_2017}
Perreault~Levasseur, L., Hezaveh, Y.~D., \& Wechsler, R.~H. 2017,
  \href{https://doi.org/10.3847/2041-8213/aa9704}{ApJ Lett., 850, L7}

\bibitem[{Petroff {et~al.}(2020)Petroff, Addison, Bennett, \&
  Weiland}]{Petroff_2020}
Petroff, M.~A., Addison, G.~E., Bennett, C.~L., \& Weiland, J.~L. 2020,
  \href{https://doi.org/10.3847/1538-4357/abb9a7}{ApJ, 903, 104}

\bibitem[{{Planck Collaboration} {et~al.}(2016){Planck Collaboration}, Ade,
  Aghanim, {et~al.}}]{Planck_2016_ISW}
{Planck Collaboration}, Ade, P. A.~R., Aghanim, N., {et~al.} 2016,
  \href{https://doi.org/10.1051/0004-6361/201525831}{A\&A, 594, A21}

\bibitem[{Repp \& Szapudi(2016)}]{Repp2016}
Repp, A. \& Szapudi, I. 2016,
  \href{https://doi.org/10.1093/mnrasl/slw178}{Monthly Notices of the Royal
  Astronomical Society: Letters, 464, L21}

\bibitem[{Ronneberger {et~al.}(2015)Ronneberger, Fischer, \&
  Brox}]{ronneberger2015}
Ronneberger, O., Fischer, P., \& Brox, T. 2015, in Lecture Notes in Computer
  Science, Vol. 9351, Medical Image Computing and Computer-Assisted
  Intervention -- MICCAI 2015, ed. N.~Navab, J.~Hornegger, W.~M. Wells, \&
  A.~F. Frangi (Cham: Springer), 234--241

\bibitem[{Villaescusa-Navarro {et~al.}(2021)Villaescusa-Navarro,
  Angl{\'e}s-Alc{\'a}zar, Genel, Spergel, Li, Wandelt, Nicola,
  {et~al.}}]{VillaescusaNavarro_2021}
Villaescusa-Navarro, F., Angl{\'e}s-Alc{\'a}zar, D., Genel, S., {et~al.} 2021,
  \href{https://doi.org/10.3847/1538-4357/abf7ba}{ApJ, 915, 71}

\bibitem[{Wright {et~al.}(2025)Wright, St{\"o}lzner, Asgari,
  {et~al.}}]{Wright_2025}
Wright, A.~H., St{\"o}lzner, B., Asgari, M., {et~al.} 2025,
  \href{https://doi.org/10.1051/0004-6361/202554908}{A\&A, 703, A158}

\end{thebibliography}

\begin{appendix}

\end{appendix}
\end{document}